\documentclass[12pt,preprint]{aastex}

\shortauthors{Wang et al.}
\begin{document}

\title{Most hard X-ray selected quasars in $Chandra$ Deep Fields are obscured}
\author{J. X. Wang\altaffilmark{1},
P. Jiang\altaffilmark{1},
Z. Y, Zheng\altaffilmark{1},
P. Tozzi\altaffilmark{2,3},
C. Norman\altaffilmark{4,9},
R. Giacconi\altaffilmark{4},
R. Gilli\altaffilmark{5},
G. Hasinger\altaffilmark{6},
L. Kewley\altaffilmark{7},
V. Mainieri\altaffilmark{6},
M. Nonino\altaffilmark{3},
P. Rosati\altaffilmark{8},
A. Streblyanska\altaffilmark{6},
G. Szokoly\altaffilmark{6},
A. Zirm\altaffilmark{4},
W. Zheng\altaffilmark{4}
}
\altaffiltext{1}{Center for Astrophysics, University of Science and Technology of China, Hefei, Anhui 230026, P. R. China; jxw@ustc.edu.cn.}
\altaffiltext{2}{INAF Osservatorio Astronomico, Via G. Tiepolo 11, 34131 Trieste, Italy}
\altaffiltext{3}{INFN - National Institute for Nuclear Physics, Trieste, Italy}
\altaffiltext{4}{Dept. of Physics and Astronomy, The Johns Hopkins University, Baltimore, MD 21218}
\altaffiltext{5}{Istituto Nazionale di Astrofisica (INAF) - Osservatorio Astrofisico di Arcetri, Largo E. Fermi 5, 50125 Firenze, Italy}
\altaffiltext{6}{Max-Planck-Institut f\"{u}r extraterrestrische Physik, Postfach 1312, D-85741 Garching, Germany}
\altaffiltext{7}{Hubble Fellow, Institute for Astronomy, University of Hawaii, 2680 Woodlawn Drive, Manoa, HI 96822}
\altaffiltext{8}{European Southern Observatory, Karl-Schwarzschild-Strasse 2, Garching, D-85748, Germany}
\altaffiltext{9}{Space Telescope Science Institute, 3700 San Martin Drive, Baltimore, MD 21218}
\vspace{0.1cm}

\begin{abstract}
Measuring the population of obscured quasars is one of the key issues to 
understand the evolution of active galactic nuclei (AGNs). 
With a redshift completeness of 99\%,
the X-ray sources detected in $Chandra$ Deep Field South (CDF-S)
provide the best sample for this issue.
In this letter we study the population of obscured quasars in CDF-S by choosing the 4 -- 7 keV selected sample,
which is less biased by the intrinsic X-ray absorption.
The 4 -- 7 keV band selected samples also filter out most of the X-ray faint
sources with too few counts, for which the measurements of
N$_H$ and L$_X$ have very large uncertainties.
Simply adopting the best-fit L$_{2-10keV}$ and N$_H$, we find 71$\pm$19\% 
(20 out of 28) of the quasars (with intrinsic L$_{2-10keV}$ $>$ 10$^{44}$
erg s$^{-1}$) 
are obscured with N$_H$ $>$ 10$^{22}$ cm$^{-2}$.
Taking account of the uncertainties in the measurements of both N$_H$ and 
L$_X$, conservative lower and upper limits of the fraction are
54\% (13 out 24) and 84\% (31 out 37).
In $Chandra$ Deep Field North, the number is 29\%, however, this is mainly due to
the redshift incompleteness. 
We estimate a fraction of $\sim$ 50\% - 63\% after correcting 
the redshift incompleteness with a straightforward approach. Our results 
robustly confirm the 
existence of a large population of obscured quasars.
\end{abstract}

\section{Introduction}
Supermassive black holes (SMBH) have been found in the center of most 
(if not all) nearby galaxies (e.g. Kormendy \& Richstone 1995; Kormendy \&
Gebhardt 2001), the growth of which is believed mainly due to the accretion of 
quasars (e.g., luminous active galactic nuclei) in the distant universe 
(Yu \& Tremaine 2002). The unified model for active galactic nuclei (AGNs;
Antonucci 1993) has predicted a large population of heavily obscured
powerful quasars called type-2 quasars. They have been predicted to be the 
same as
type-1 quasars but with strong obscuration in both optical and soft X-ray band.
Most of such type-2 quasars, which might dominate the black hole growth
(e.g., Mart\'\i nez-Sansigre et al. 2005), have been missed by optical surveys for quasars.

The hard X-ray emission is less biased by the obscuration, making hard X-ray
surveys a good approach to search for type-2 quasars. Recent 
deep and 
wide-area X-ray surveys performed by $Chandra$ and XMM have revealed a number 
of such sources (Norman et al. 2002; Stern et al. 2002; Mainieri et al. 2002; Fiore et al. 2003; Caccianiga et al. 2004).
Using various $Chandra$
and $ASCA$ surveys, Steffen et al. (2003) claimed that 
broad-line sources dominate above $L_{2 - 8 keV}$ = 3 $\times$ 10$^{43}$ ergs s$^{-1}$, and type II AGNs become important only at Seyfert-like X-ray 
luminosities. Similarly, Ueda et al. (2003) computed the X-ray luminosity 
function for 2 -- 10 keV selected AGN samples, and found that the fraction of
X-ray absorbed AGNs (with $N_H$ $>$ 10$^{22}$ cm$^{-2}$) drops from $\sim$
0.6 at intrinsic $L_X$ around 10$^{42}$ ergs s$^{-1}$ to around 0.3
at $L_X$ above 10$^{44}$ ergs s$^{-1}$.
Several other studies based on hard X-ray surveys also
support the scheme that the fraction of type II AGNs (or X-ray absorbed
AGNs) decreases with intrinsic luminosity (Barger et al. 2005; 
La Franca et al. 2005).
However, contrary results are also reported. By performing Monte-Carlo 
simulations to match the X-ray colors in the 13$^H$ {\it XMM-Newton}
deep field, Dwelly et al. (2005) claimed that the fraction of obscured AGN depends on neither the luminosity nor the redshift. Perola et al. (2004) found that
the fraction of obscured AGNs in HELLAS2XMM does not change with X-ray 
luminosity, although the fraction (40\%) appears smaller than expected.
Also see Eckart et al. (2006) and Dwelly \& Page (2006) for most recent
works.
Note that the observed fraction of obscured sources is a function
of the X-ray flux due to the bias of the absorption and/or its evolution
with redshift (see e.g. Comastri et al. 2001; Piconcelli
et al. 2002, 2003; Ueda et al. 2003; La Franca et al. 2005),
therefore one should be cautious to compare results from surveys with
different depths. 

The 2 Ms $Chandra$ exposure on $Chandra$ Deep Field North (CDF-N, Brandt et al. 2001; Alexander et al. 2003) and 1 Ms exposure on $Chandra$ Deep Field South (CDF-S,
Giacconi et al. 2002) are the deepest X-ray images ever taken.
Tozzi et al. (2006, hereafter T06) presented the detailed X-ray spectral
studies and obtained absorption corrected luminosities (which is essential
for the identification of quasars) for the X-ray sources in CDF-S.
In this letter, we present samples of 4 -- 7 keV band selected AGNs in 
$Chandra$ Deep Fields, and employed such samples to study the population 
of obscured quasars. 
We point out that comparing with the normally used hard band ( 2 -- 7/8/10 keV),
4 -- 7 keV band samples are less biased against X-ray photo-electronic 
absorption, therefore are better suited to study the fraction of obscured 
sources.
Throughout this paper, H0 is taken to be 70 ks s$^{-1}$ Mpc$^{-1}$, 
$\Omega_m$=0.3 and $\Omega_{\lambda}$=0.7 (Spergel et al. 2003).

\section{The X-ray Data Reduction}
The 1 Ms $Chandra$ exposure on the CDF-S was composed of 
eleven individual ACIS observations obtained from
October 1999 to December 2000.
The detailed data reduction and analysis are described in
Giacconi et al. (2001; 2002), Tozzi et al. (2001) and
Rosati et al. (2002). An X-ray catalog of 347 X-ray sources
was presented in Giacconi et al. (2002). 
The 2 Ms $Chandra$ exposure on the CDF-N was composed of
twenty individual ACIS observations obtained from
November 1999 to February 2002 (Alexander et al. 2003). An X-ray catalog of 503 X-ray sources
was presented in Alexander et al. (2003). 

In this letter we use updated X-ray data reductions with CIAO3.2.2 and 
CALDB3.1.0 on both CDF-S and CDF-N, therefore including the most updated 
corrections. For each individual observation, the ACIS hot pixels and 
cosmic ray afterglows were re-identified using the new CIAO script
``acis$\_$run$\_$hotpix".
The level 1 data were then reprocessed to clean the ACIS particle background
for both FAINT and VFAINT mode observations, and filtered to include
only the standard event grades 0,2,3,4,6. The recently released
time-dependent gain correction and ACIS charge transfer inefficiency (CTI)
correction were also applied.
Due to the large off-axis angle during the observations,
the ACIS-S chips have poorer spatial
resolution and effective area than the ACIS-I chips.
In this paper, data from any ACIS-S CCD are ignored.
All bad pixels and columns
were also removed. High background time intervals were finally removed
from level 2 files. 
The offsets between the astrometry of individual observations
were obtained by registering the X-ray sources showing up in both exposures.
4 -- 7 keV band X-ray images were extracted from the combined event files
for the two fields.

We run WAVDETECT (Freeman et al. 2002) on the extracted 4 -- 7 keV
band X-ray images using a probability threshold of 1 $\times$ 10$^{-7}$ 
(corresponding
to 0.5 false sources expected per image), and wavelet scales of
1, $\sqrt{2}$, 2, 2$\sqrt{2}$, 4, 4$\sqrt{2}$, 8, 8$\sqrt{2}$, 16 pixels 
(1 pixel = 0.492\arcsec).
A total of 107 X-ray sources were detected in 4 -- 7 keV band in CDF-S
and 176 in CDF-N. We state that no new sources were detected
in additional to the published catalogs.
Rosati et al. (2002) stated that 110 X-ray sources in CDF-S show
S/N $>$ 2.1 in 5 -- 7 keV band, however, these sources are pre-selected
by running SExtractor on the 0.5 -- 7 keV band image and filtered with X-ray 
photometry measurements, thus are not selected independently in 
the 5 -- 7 keV band and are biased by softer band brighter sources which 
have higher chance to be pre-selected by SExtractor. 
By comparing with the catalogs in Szokoly et al. (2004) and Zheng et al.
(2004), we got the spectroscopic/photometric redshifts for all the 107 sources 
in CDF-S. 54 out of them have secure spectroscopic redshifts, and for sources
with photometric redshifts, a medium uncertainty (at 95\% confidence level)
of 0.17 in the redshift is expected. Considering that we are studying quasars
which are located at much higher redshifts, the uncertainties in the redshift
won't significantly affect our main results in this paper.
We also make a quick estimation on the reliability of the optical 
counterparts. Zheng et al. (2004) provides the offsets of counterparts from X-ray
positions. Taking the offsets as radius of circles, the total area within the
circles are 918 square arc seconds for our 107 X-ray sources, and there are at 
most 7 random sources (down to R = 26, assuming a density of 100,000 per square
degree) in such a region. Considering that most of X-ray sources have optical 
counterparts much brighter than R=26 which could outshine fainter spurious 
sources, and a large fraction of them have been confirmed to be AGN by optical 
spectra, the true number of spurious optical counterparts within our 107 
sources is much smaller than 7, thus will not either affect our main results 
presented in this paper.
 
\begin{figure}
\plotone{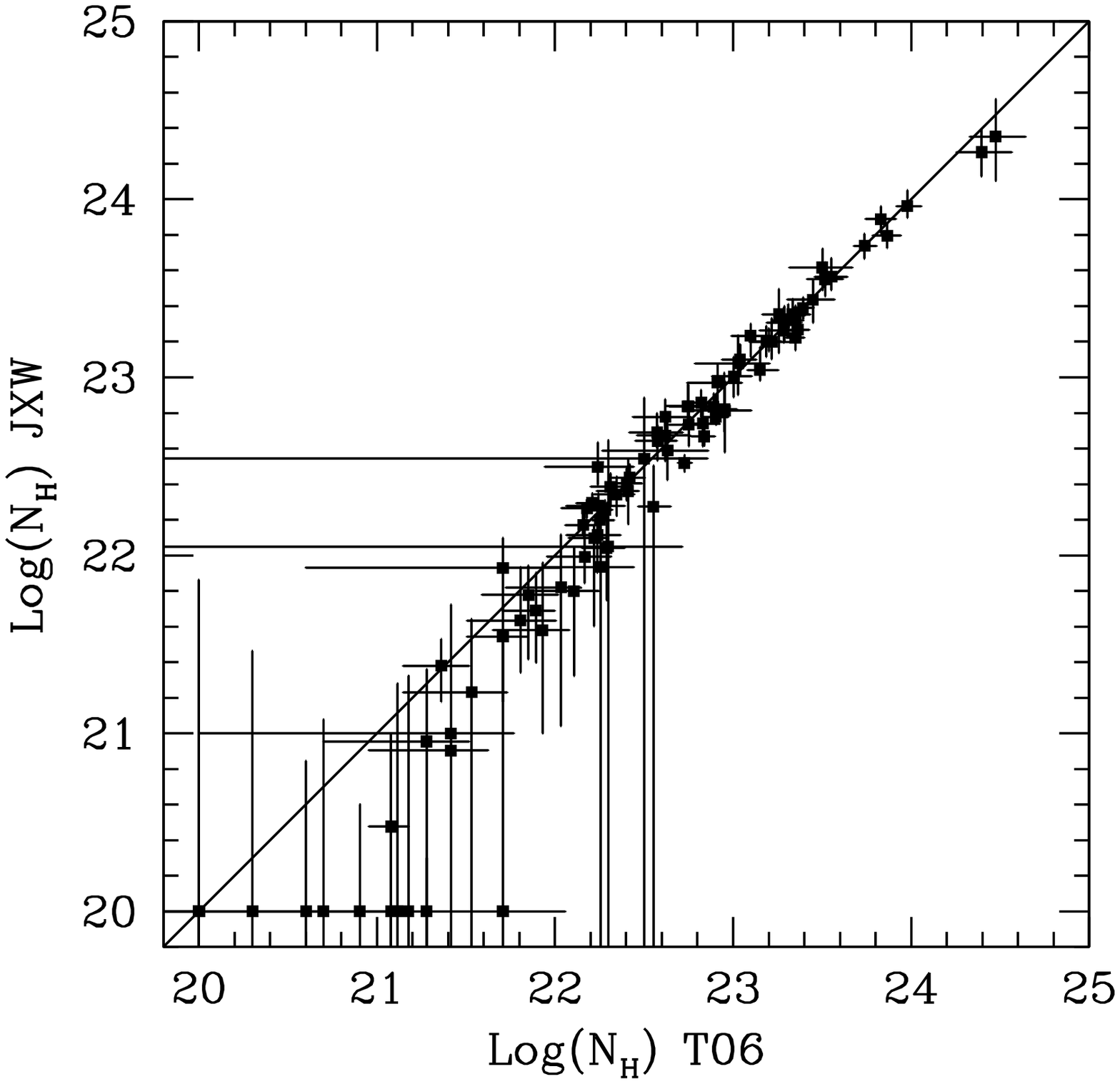}
\caption{
The best-fit absorption column densities (with 1$\sigma$ errorbars) from X-ray spectra with
different version of Chandra calibrations. The X axis plots N$_H$ from
T06 (with CALDB2.26) and the Y axis plots the independently derived N$_H$
from re-calibrated spectra (with CALDB3.1.0). For display purpose, we plot
best-fit N$_H$=0 at N$_H$ = 10$^{22}$ cm$^{-2}$.
For faint sources in T06 for which $\Gamma$ was fixed at 1.8, we adopted
the same fixed $\Gamma$ in this figure.
}
\label{NH}
\end{figure}

Following T06
we perform X-ray spectral fitting to obtain their X-ray absorption column density, and absorption corrected
rest frame 2 -- 10 keV luminosities.
The source spectrum was extracted from a circle of radius $R_s$ = 2.4 $\times$ $FWHM$ and the background extracted from an annulus with outer radius $R_s$+12\arcsec and inner radius $R_s$+2\arcsec, after masking out other sources.
XSPEC11.3.1 and Cash statistics were adopted to perform the spectral fits
in 0.6 -- 7.0 keV band.
We found that our own fitting results are consistent (within the
fitting uncertainty, see Fig. \ref{NH}) with 
T06 which used an earlier version of 
$Chandra$ calibration CALDB2.26 and CIAO3.0.1. 
This indicates that the uncertainty in the
data calibration does not significantly affect our results presented in this
paper.
For the 176 X-ray sources in CDF-N, we got 127 spectroscopic/photometric
redshifts from Barger et al. (2003), yielding a redshift completeness of
72\%. 

\section {Why 4 -- 7 keV?}
It's well known that harder band X-ray emission is less affected by 
photo-electronic absorption, thus the harder X-ray band is less biased
against absorption, and is best suited to select obscured sources
(Fiore et al. 1999; Comastri et al. 2001; Nandra et al. 2003)\footnote{
Note this is not true for Compton-thick sources, since for Compton-thick 
absorption, even $\gamma$-ray emission could be strongly biased by Compton 
scattering, see Wang \& Jiang 2006).}.
In this paper, we choose to adopt the 4 -- 7 keV selected sample to
study the population of obscured quasars in CDFs. Since 4 -- 7 keV band
is less biased against the absorption than the traditional 2 -- 7/8 keV band, 
the observed fraction of obscured quasars in this band is less biased and
is expected to be closer to the true value.

\begin{figure}
\plotone{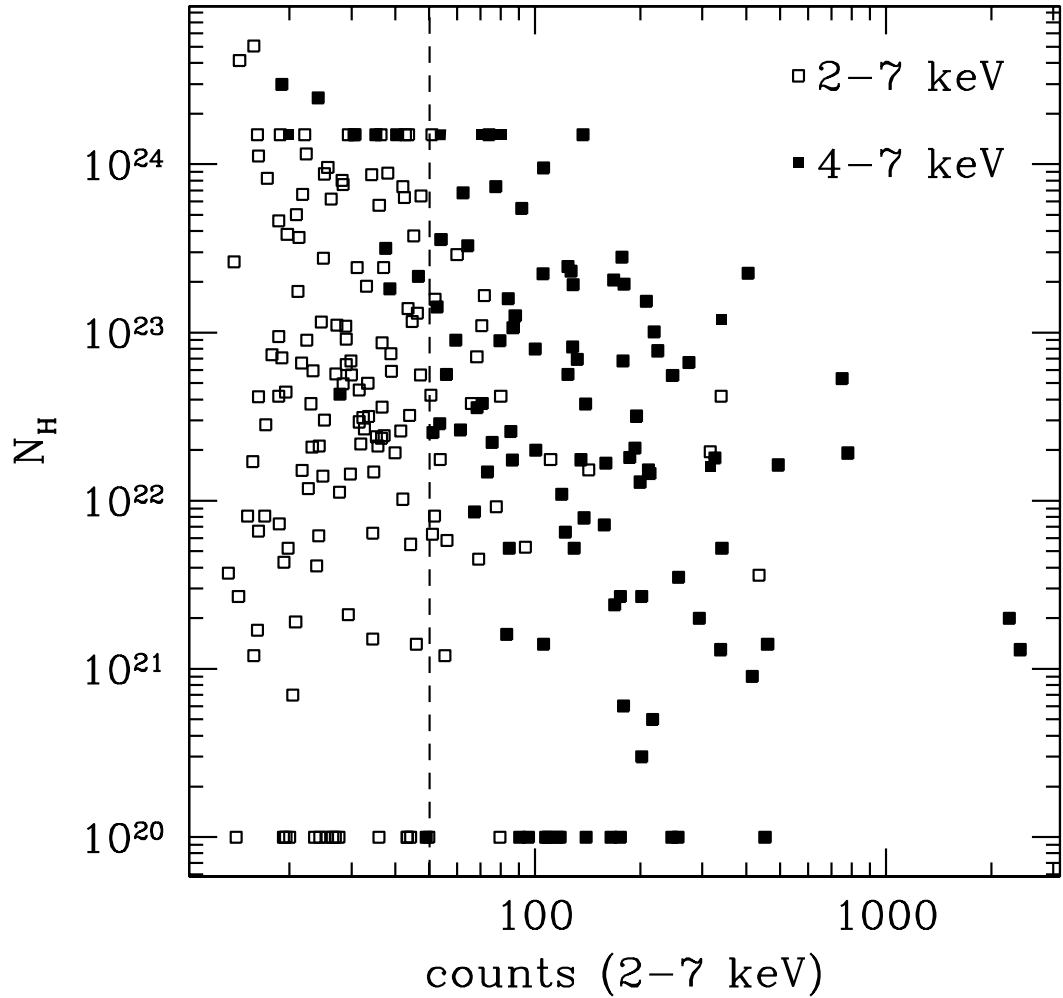}
\caption{The X-ray absorption column density N$_H$ of the 2 -- 7 keV selected
AGNs in CDF-S versus the X-ray counts in 2 -- 7 keV detected by $Chandra$
(open squares). The 4 -- 7 keV detected sources are overplotted as solid
squares. To the right of the dashed vertical line, the total number of sources
is 107, the same as that of the 4 -- 7 keV selected sources.
}
\label{counts}
\end{figure}

Another significant advantage of using 4 -- 7 keV band sample is to filter out
faint sources with too few X-ray counts, for which the X-ray spectral fitting
yield too large uncertainties and could produce artificial high absorption
and luminosity (especially at high z, see simulations in T06).
In Fig. \ref{counts} we plot N$_H$ versus 2 -- 7 keV band counts for the 2 -- 7 keV selected sources in CDF-S (open squares). The 107 4 -- 7 keV band
detected sources are overplotted as solid squares.
Comparing with a 2 -- 7 keV counts limited bright sample with the same number of sources,
the 4 -- 7 keV band sample picked up more sources with N$_H$ $\gtrsim$
10$^{23}$ cm$^{-2}$, thus is clearly more complete to obscured sources.
The 4 -- 7 keV band sample also filters out most of
faint X-ray sources with very few counts, making it more suitable for
X-ray spectral analysis.

\begin{figure}
\plottwo{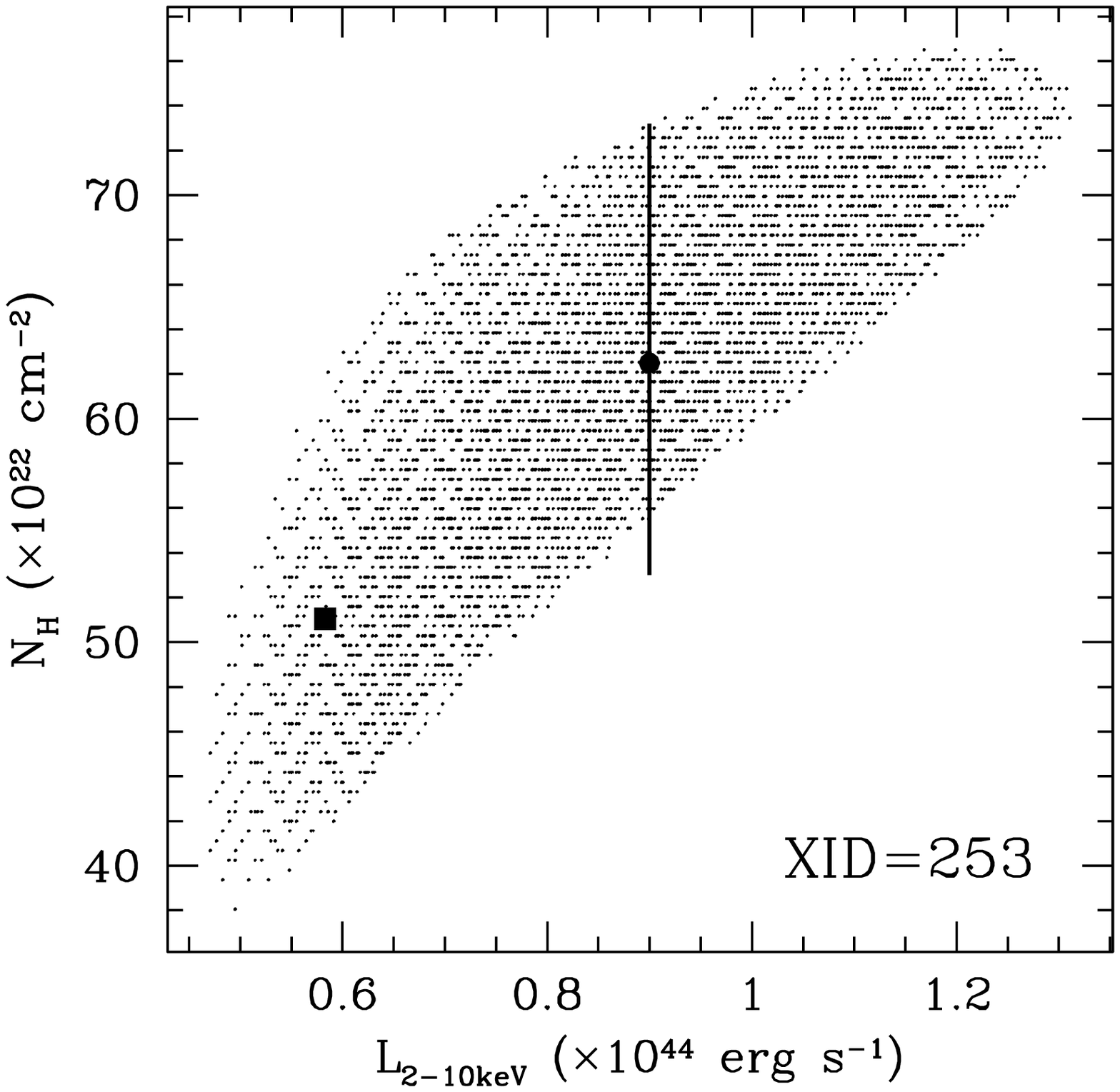}{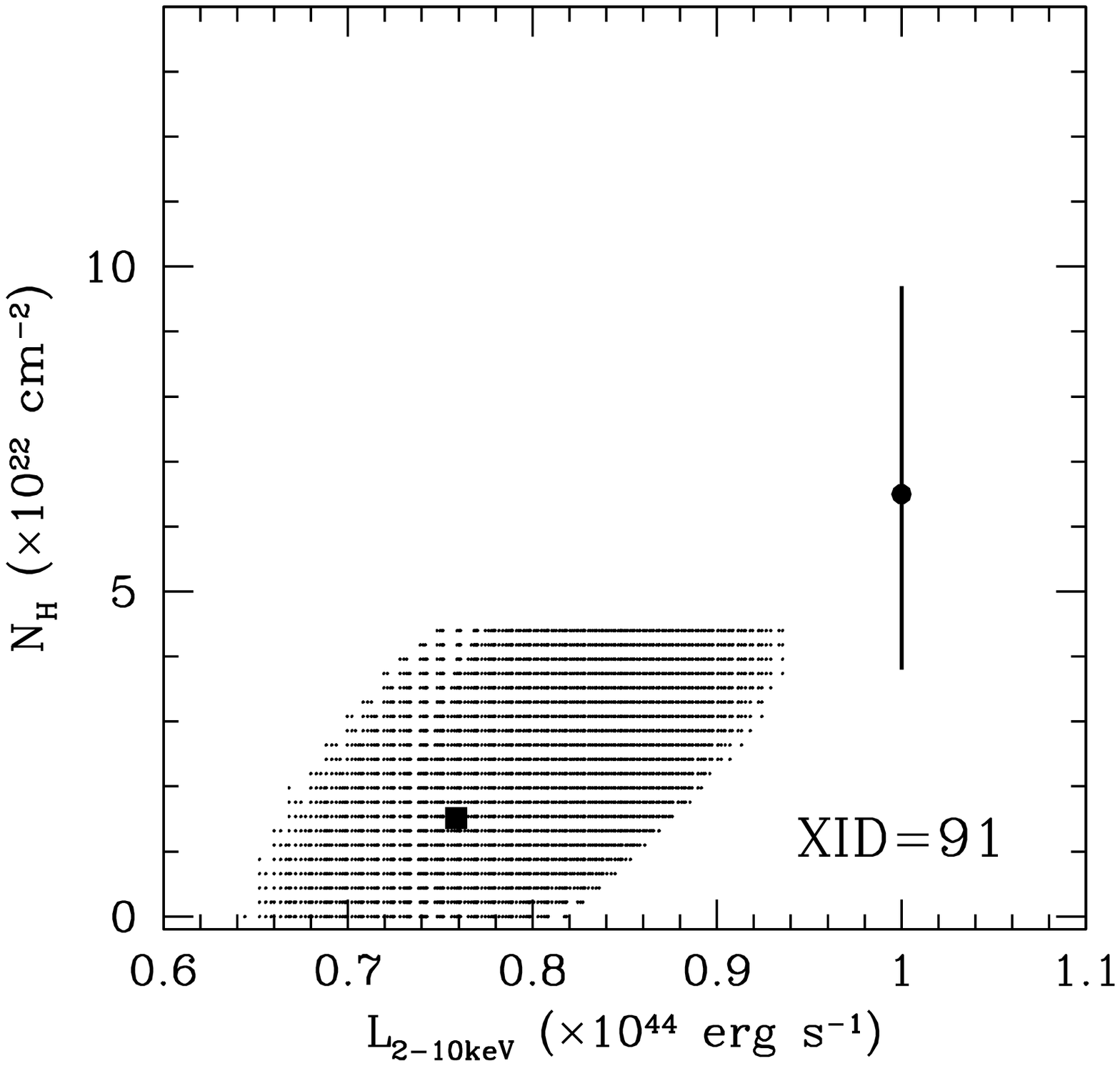}
\caption{
The 90\% confidence range ($\Delta$C=2.706) of N$_H$ and L$_X$ for two 
possible obscured quasars. The confidence range was calculated by setting 
the photon index $\Gamma$ varying from 1.4 to 2.4  with best-fit N$_H$ and 
L$_X$ marked as black squares.
The best-fit values with $\Gamma$ fixed at 1.8 are also plotted as
black dots with 90\% errorbars for N$_H$ (solid line). 
}
\label{samples}
\end{figure}

\begin{figure}
\plotone{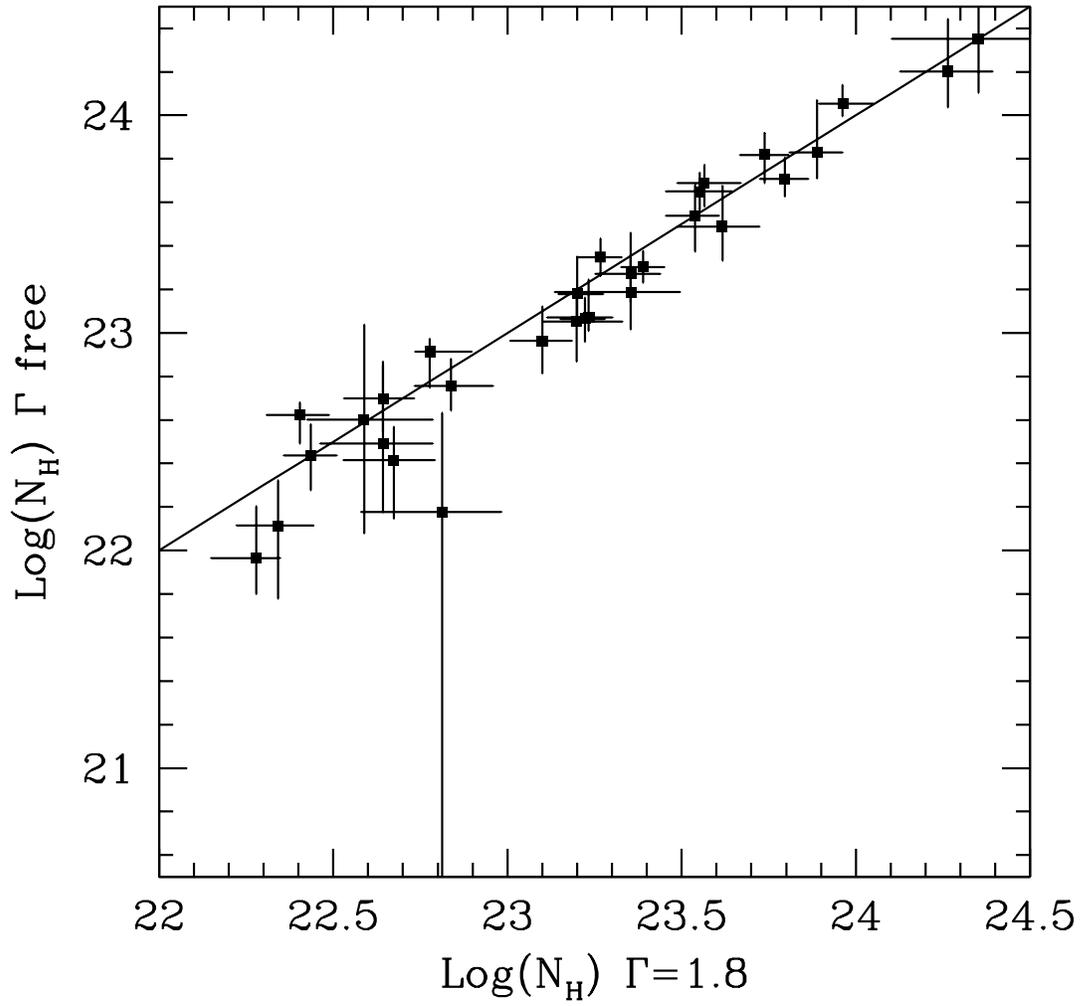}
\caption{
The best-fit absorption column density (with 1$\sigma$ errorbars) with 
$\Gamma$ free versus that with $\Gamma$ fixed at 1.8 for faint sources 
in 4 -- 7 keV band sample. The solid line plots X = Y.
}
\label{NH2}
\end{figure}

During the spectral fit, we do not fix the photon index $\Gamma$ at 1.8, 
but allow it vary from 1.4 to 2.4 for faint sources (since
too low or too high values might not be physical). This is because a) while an average $\Gamma$
= 1.8 is valid for an AGN sample, the fitting results for individual
sources could be strongly biased due to the scattering of $\Gamma$;
b) the uncertainties in the measurements could be significantly underestimated
by fixing $\Gamma$. The fitting results are shown in table 1.
We also point out that the measurements of the intrinsic L$_X$ also
have large uncertainties, especially for faint and obscured sources.
In Fig. \ref{samples} we plot the 90\%
confidence region ($\Delta$C = 2.706) of N$_H$ and L$_X$ for two
faint sources (which would be classified as obscured quasars in T06).
Instead of fixing the photon index at 1.8, the confidence region was
calculated by setting the photon index free (varying from 1.4 to 2.4).
In the figure we can clearly see the large uncertainties in the
measurement of L$_X$ and N$_H$ which both should also be taken into account
while classifying obscured quasars. We can also see the potential bias caused
to the measurement of N$_H$ and L$_X$ by fixing $\Gamma$ at 1.8.
In Fig. \ref{NH2} we plot the best fit N$_H$ obtained with $\Gamma$ free versus
that with $\Gamma$ fixed at 1.8. We can clearly see that fixing $\Gamma$ at 1.8
introduces obvious scattering for individual sources in the measurement of 
N$_H$, although there is no systematic difference between two values. Fixing $\Gamma$
also significantly underestimate the uncertainty of N$_H$, which would also
lead to obvious uncertainty in the measurement of L$_X$ (see Fig. \ref{samples}).

\section{The fraction of obscured quasars}
\begin{figure}
\plotone{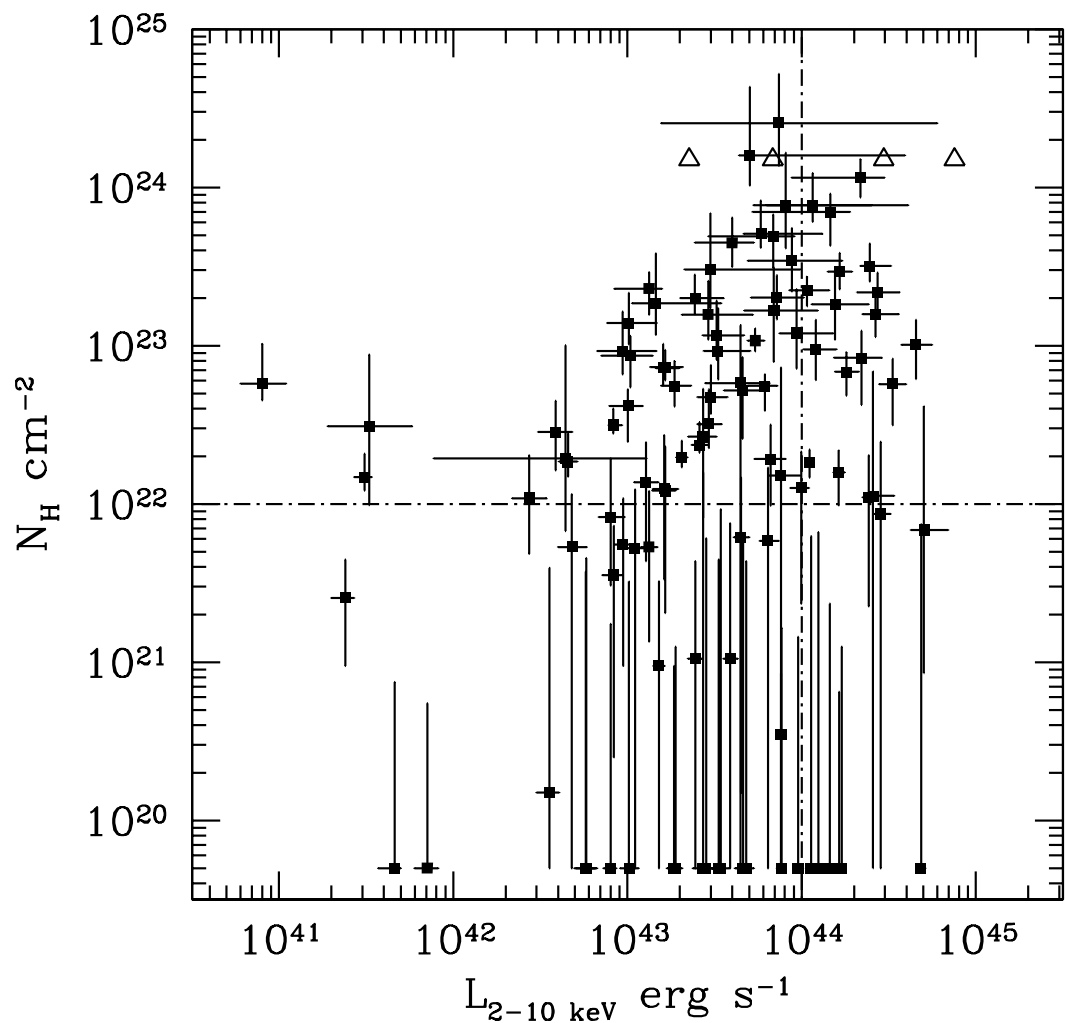}
\caption{
Our best-fit absorption column density versus absorption corrected 
L$_{2-10keV}$ with 90\% errorbars for 107 4 -- 7 keV band selected
sources in CDF-S. All data are from table 1. Triangles are sources
fitted with reflection model ($pexrav$, see table 1 for details).
For display purpose, we assign 0.5$\times$10$^{20}$ cm$^{-2}$ to
sources with best-fit N$_H$ at zero. Our selection criteria for
obscured quasars are plotted as dash-dotted lines.
}
\label{107}
\end{figure}

The output N$_H$ versus L$_X$ for 107 sources selected in 4 -- 7 keV band in
CDF-S is presented in Fig. \ref{107}.
Our selection criteria for obscured quasars (L$_{2-10keV}$ $>$
10$^{44}$ erg s$^{-1}$ and N$_H$ $>$ 10$^{22}$ cm$^{-2}$) are also plotted.
Among the 107 sources,
we identify 13 secure obscured quasars 
(XID 18,25,27,35,45,57,62,68,72,76,153,159,202)
with both L$_{2-10keV}$ $>$ 
10$^{44}$ erg s$^{-1}$ and N$_H$ $>$ 10$^{22}$ cm$^{-2}$ at $>$ 90\% 
confidence level and 6 secure unobscured quasars
(XID 11,22,42,60,67,206). There are 5 more
quasars (XID 6,7,24,31,61) with L$_{2-10keV}$ $>$
10$^{44}$ erg s$^{-1}$ at $>$ 90\% confidence level but 90\% N$_H$
uncertainty range across 10$^{22}$ cm$^{-2}$. 
A conservative lower limit in the fraction of X-ray obscured quasars
(with N$_H$ $>$ 10$^{22}$ cm$^{-2}$) is thus 54\%. 
Considering there are 13 possible obscured quasars with 90\% L$_{2-10keV}$
uncertainty range across 10$^{44}$ erg s$^{-1}$ (12 with N$_H$ $>$ 10$^{22}$ 
cm$^{-2}$ at $>$ 90\% confidence level, XID 51,54,152,156,209,227,243,253,259,263,543,601,609), the upper limit of the fraction
could be 84\%.
We can clearly see that due to the large uncertainties in the measurement
of both N$_H$ and L$_{2-10keV}$, the fraction of obscured quasars in CDF-S
ranges from 54\% to 84\%. If we simply adopt the best-fit N$_H$ and 
L$_{2-10keV}$, the fraction is 71\% (20 out 28).
Apply the same selection criteria to table 1 in T06 (with best-fit
N$_H$ $>$ 10$^{22}$ and L$_{2-10keV}$ $>$ 10$^{44}$ erg s$^{-1}$),
we obtain a slightly higher fraction of 85\% (39 out 46, but 
consistent within 1$\sigma$ uncertainty). 
Considering that most of our 4 -- 7 keV band selected sources are among
the brightest in the CDF-S sample (see Fig. \ref{counts}), and since the absorbed
fraction is increasing at lower fluxes, two values well agree with
each other.

Among the 13 secure obscured quasars and 18 extra possible obscured quasars,
we find optical classification for 19 of them from Szokoly et al. (2004).
Only three (XID 24,62,68) are classified as broad line AGN, and all of them
show associated absorption systems in their optical spectra (XID 62 is
identified as BAL QSO), suggesting an outflow origin for their X-ray 
obscuration. 
This indicates that most of our X-ray
obscured quasars are likely type II QSO with broad line region also obscured.

\begin{figure}
\plotone{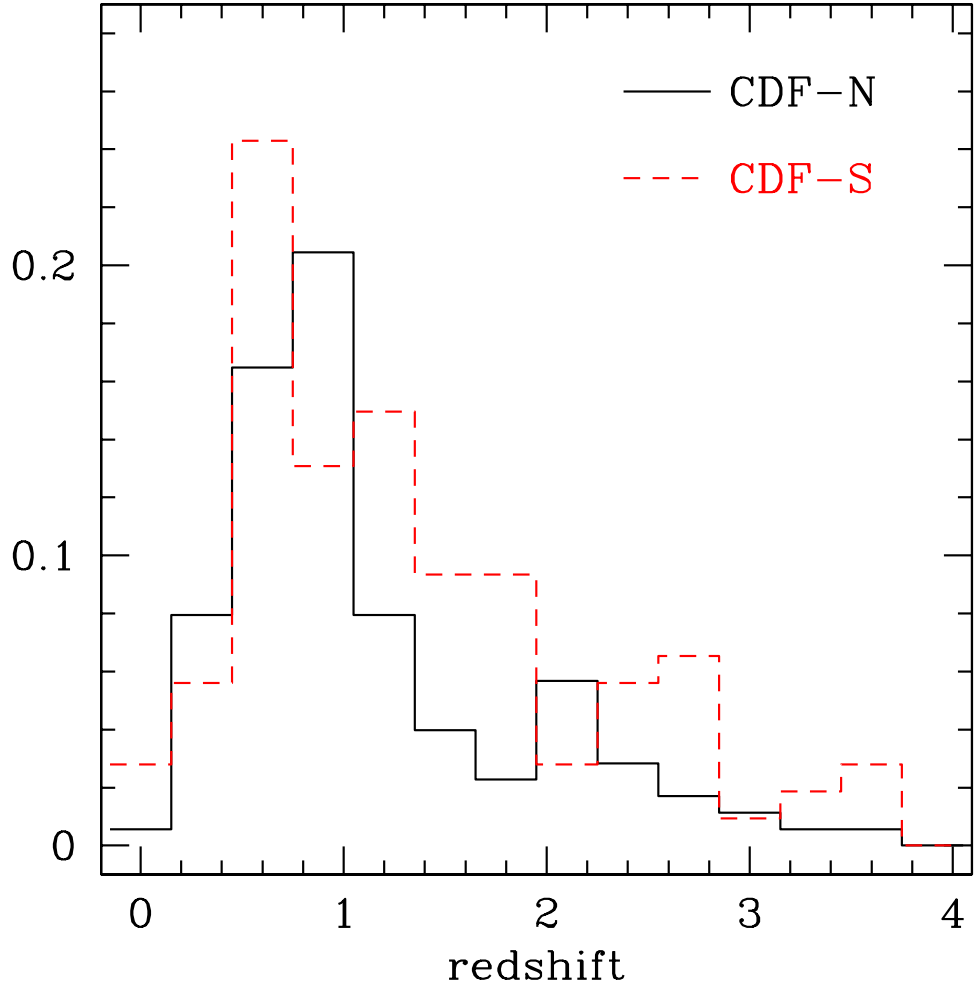}
\caption{The fractional redshift distribution of our 4 -- 7 keV band selected 
sources in CDF-S (dashed line) and CDF-N (solid line). Note the redshift 
completeness for CDF-S sample is 100\% and 72\% for CDF-N sample. 
}
\label{cdfsnz}
\end{figure}

In CDF-N, we obtained 21 quasars with best-fit L$_{2-10keV}$ $>$ 10$^{44}$ 
erg s$^{-1}$, 6 of which have best-fit N$_H$ $>$ 10$^{22}$ cm$^{-2}$, corresponding 
to a much smaller fraction of 29\%. 
However we note that the total number of quasars in CDF-N (21) is much less 
than that in CDF-S (28) although CDF-N is twice as deep and is expected to 
detect more quasars. This is a consequence of the fact that a significant 
fraction of the quasars 
have no redshift available. In Fig. \ref{cdfsnz} we plot the redshift 
distribution of the 4 -- 7 keV band sources in CDF-S and CDF-N. Assuming that 
the CDF-N sample has similar redshift distribution to that of CDF-S sample, 
we can see in the figure that most of the sources without redshifts in CDF-N 
should locate at redshift between 1 and 4. 
Assuming all sources without redshifts in CDF-N are located at z = 2 (or 3),
we perform spectral fitting to estimate their N$_H$ and L$_{2-10keV}$. After
such a coarse correction to redshift incompleteness, we find 
the fraction of obscured quasars could increase to 50\% (or 63\%)
and the total number of quasars could increase to 31 (or 41).

\section{Conclusions}
We present 4 -- 7 keV band selected sources in CDF-S.
We find that 54\% -- 84\% of the 4 -- 7 keV band quasars in CDF-S are 
obscured with N$_H$ $>$ 10$^{22}$ cm$^{-2}$. This result is well consistent
with that of Dwelly \& Page (2006) who find that $\sim$ 75\% XMM-$Newton$
sources in CDF-S are obscured at all luminosities.
We also note that a most recent work by Georgantopoulo, Georgakakis \& Akylas 
(2006) obtained a similar
fraction of X-ray obscured quasars in CDF-S (74\%, 17 out of 23) by fitting
X-ray spectra of 186 2 -- 10 keV band detected sources covered by all 11
$Chandra$ pointings.
We indicate that using different versions of $Chandra$ calibrations
yield consistent fitting results, suggesting that calibration uncertainty
would not significantly affect our main results presented in this paper.
The uncertainties in the measurement of both L$_X$ and N$_H$ are also
carefully taken into account. 
We note that AGN's X-ray spectra are often more complicated than an absorbed 
powerlaw model (T06), especially in the soft band (such as warm absorber
and soft excess), however we note that most of the quasars are located
at high redshift ($z$ $>$ 1) for which extra soft features would have
been shifted out of $Chandra$ bandpass. 
In CDF-N, the fraction is much lower (29\%), however, we
found this is probably due to the redshift incompleteness in the CDF-N sample.
After correcting the redshift completeness, we found that 50\%-63\% of the
quasars in CDF-N are obscured, consistent with that in CDF-S (54\% - 84\%).
Our results
robustly confirm the
existence of a large population of obscured quasars.

\acknowledgments
The work of JW was supported by Chinese NSF through NSFC10473009, NSFC10533050 and the CAS "Bai Ren" project at University of Science and Technology of China.
PT acknowledges financial contribution from contract ASI--INAF I/023/05/0.

\clearpage

\begin{deluxetable}{lcccccc}
\tabletypesize{\scriptsize}
\tablewidth{0pc}
\tablecaption{X-ray spectral fits to 4 -- 7 keV selected sources in CDF-S}
\tablehead{
\colhead{XID} & \colhead{z} & \colhead{Q} & \colhead{N$_H$/10$^{22}$cm$^{-2}$} & \colhead{$\Gamma$} &  \colhead{L$_{2-10keV}$ erg s$^{-1}$} 
}
\startdata
22 & 1.92 & 3.0 & 	  0.00$^{+  0.66}_{-  0.00}$ &	  1.81$^{+  0.16}_{-  0.11}$ &	  1.25$_{-  0.08}^{+  0.08}\times10^{44}$\\
27 & 3.06 & 3.0 & 	 31.85$^{+ 12.20}_{-  6.33}$ &	  1.40$^{+  0.17}_{-  0.00}$ &	  2.45$_{-  0.28}^{+  0.81}\times10^{44}$\\
39 & 1.22 & 3.0 & 	  0.00$^{+  0.14}_{-  0.00}$ &	  1.78$^{+  0.07}_{-  0.08}$ &	  9.54$_{-  0.47}^{+  0.41}\times10^{43}$\\
202 & 3.70 & 3.0 & 	 150 &	  2.12$^{+  0.25}_{-  0.22}$ &	  7.53$\times10^{44}$\\
151 & 0.60 & 3.0 & 	 22.94$^{+  6.26}_{-  7.21}$ &	  2.40$^{+  0.00}_{-  0.77}$ &	  1.33$_{-  0.49}^{+  0.25}\times10^{43}$\\
153 & 1.54 & 3.0 & 	 150 &	  1.55$^{+  0.26}_{-  0.15}$ &	  2.69$\times10^{44}$\\
63 & 0.54 & 3.0 & 	  0.03$^{+  0.05}_{-  0.03}$ &	  1.86$^{+  0.06}_{-  0.04}$ &	  7.60$_{-  0.25}^{+  0.15}\times10^{43}$\\
256 & 1.53 & 0.5 & 	 49.24$^{+ 18.16}_{- 18.31}$ &	  2.40$^{+  0.00}_{-  0.93}$ &	  6.88$_{-  3.98}^{+  2.23}\times10^{43}$\\
26 & 1.65 & 0.5 & 	  5.20$^{+  3.24}_{-  2.61}$ &	  1.95$^{+  0.45}_{-  0.42}$ &	  4.57$_{-  0.99}^{+  1.54}\times10^{43}$\\
79 & 1.82 & 0.5 & 	  0.00$^{+  1.58}_{-  0.00}$ &	  1.71$^{+  0.42}_{-  0.27}$ &	  2.71$_{-  0.35}^{+  0.58}\times10^{43}$\\
31 & 1.60 & 3.0 & 	  1.58$^{+  0.61}_{-  0.61}$ &	  2.10$^{+  0.14}_{-  0.16}$ &	  1.63$_{-  0.11}^{+  0.13}\times10^{44}$\\
36 & 1.03 & 0.5 & 	  1.37$^{+  1.08}_{-  0.94}$ &	  1.99$^{+  0.41}_{-  0.37}$ &	  1.27$_{-  0.19}^{+  0.23}\times10^{43}$\\
38 & 0.74 & 3.0 & 	  0.00$^{+  0.09}_{-  0.00}$ &	  1.94$^{+  0.10}_{-  0.10}$ &	  1.84$_{-  0.13}^{+  0.14}\times10^{43}$\\
42 & 0.73 & 3.0 & 	  0.00$^{+  0.06}_{-  0.00}$ &	  1.93$^{+  0.05}_{-  0.04}$ &	  1.64$_{-  0.03}^{+  0.05}\times10^{44}$\\
145 & 1.50 & 0.5 & 	 11.57$^{+  7.69}_{-  3.45}$ &	  1.40$^{+  0.50}_{-  0.00}$ &	  3.25$_{-  0.56}^{+  1.43}\times10^{43}$\\
51 & 1.10 & 3.0 & 	 22.34$^{+  5.02}_{-  4.58}$ &	  1.77$^{+  0.38}_{-  0.37}$ &	  1.07$_{-  0.23}^{+  0.37}\times10^{44}$\\
53 & 0.68 & 3.0 & 	  0.00$^{+  0.37}_{-  0.00}$ &	  1.63$^{+  0.29}_{-  0.17}$ &	  5.75$_{-  0.70}^{+  0.80}\times10^{42}$\\
55 & 0.12 & 3.0 & 	  1.47$^{+  0.60}_{-  0.26}$ &	  1.43$^{+  0.39}_{-  0.03}$ &	  3.10$_{-  0.40}^{+  0.30}\times10^{41}$\\
257 & 0.55 & 1.5 & 	  150 &	  1.89$^{+  0.49}_{-  0.49}$ &	  2.26$\times10^{43}$\\
62 & 2.81 & 3.0 & 	 21.81$^{+  7.05}_{-  6.50}$ &	  1.81$^{+  0.30}_{-  0.29}$ &	  2.73$_{-  0.66}^{+  0.93}\times10^{44}$\\
64 & 0.13 & 0.4 & 	  0.25$^{+  0.19}_{-  0.16}$ &	  1.74$^{+  0.27}_{-  0.23}$ &	  2.40$_{-  0.40}^{+  0.30}\times10^{41}$\\
259 & 1.76 & 0.5 & 	 69.81$^{+ 21.58}_{- 26.97}$ &	  2.33$^{+  0.07}_{-  0.88}$ &	  1.46$_{-  0.94}^{+  0.44}\times10^{44}$\\
18 & 0.98 & 3.0 & 	  1.82$^{+  0.38}_{-  0.36}$ &	  1.77$^{+  0.13}_{-  0.12}$ &	  1.11$_{-  0.06}^{+  0.06}\times10^{44}$\\
147 & 0.99 & 0.5 & 	 19.97$^{+  8.17}_{-  4.13}$ &	  1.40$^{+  0.58}_{-  0.00}$ &	  2.44$_{-  0.43}^{+  1.12}\times10^{43}$\\
20 & 1.02 & 3.0 & 	  5.56$^{+  2.43}_{-  1.44}$ &	  1.60$^{+  0.44}_{-  0.20}$ &	  1.87$_{-  0.30}^{+  0.45}\times10^{43}$\\
28 & 1.22 & 3.0 & 	  1.24$^{+  1.48}_{-  0.91}$ &	  1.46$^{+  0.41}_{-  0.06}$ &	  1.63$_{-  0.23}^{+  0.33}\times10^{43}$\\
33 & 0.67 & 3.0 & 	  0.09$^{+  0.23}_{-  0.09}$ &	  1.59$^{+  0.16}_{-  0.13}$ &	  1.51$_{-  0.11}^{+  0.12}\times10^{43}$\\
148 & 1.74 & 0.5 & 	  9.23$^{+  7.92}_{-  3.11}$ &	  1.47$^{+  0.58}_{-  0.07}$ &	  3.30$_{-  0.57}^{+  1.78}\times10^{43}$\\
515 & 2.28 & 0.5 & 	 30.20$^{+ 39.02}_{- 12.96}$ &	  1.40$^{+  0.95}_{-  0.00}$ &	  3.00$_{-  0.87}^{+  6.87}\times10^{43}$\\
78 & 0.96 & 3.0 & 	  0.00$^{+  0.32}_{-  0.00}$ &	  2.05$^{+  0.24}_{-  0.19}$ &	  1.02$_{-  0.11}^{+  0.14}\times10^{43}$\\
81 & 2.59 & 0.5 & 	  5.81$^{+  7.69}_{-  5.81}$ &	  2.03$^{+  0.37}_{-  0.62}$ &	  4.44$_{-  1.64}^{+  2.09}\times10^{43}$\\
253 & 1.89 & 1.9 & 	 51.03$^{+ 31.53}_{-  9.41}$ &	  1.40$^{+  0.76}_{-  0.00}$ &	  5.83$_{-  1.20}^{+  7.29}\times10^{43}$\\
254 & 0.10 & 0.7 & 	  5.73$^{+  4.52}_{-  1.22}$ &	  1.40$^{+  0.89}_{-  0.00}$ &	  8.00$_{-  2.00}^{+  3.00}\times10^{40}$\\
50 & 0.67 & 1.9 & 	  1.08$^{+  0.94}_{-  0.60}$ &	  1.40$^{+  0.18}_{-  0.00}$ &	  2.73$_{-  0.55}^{+  0.70}\times10^{42}$\\
52 & 0.57 & 3.0 & 	  0.00$^{+  0.17}_{-  0.00}$ &	  1.96$^{+  0.17}_{-  0.10}$ &	  7.99$_{-  0.73}^{+  0.66}\times10^{42}$\\
56 & 0.60 & 3.0 & 	  1.96$^{+  0.55}_{-  0.26}$ &	  1.40$^{+  0.19}_{-  0.00}$ &	  2.05$_{-  0.14}^{+  0.17}\times10^{43}$\\
60 & 1.61 & 3.0 & 	  0.00$^{+  0.23}_{-  0.00}$ &	  1.88$^{+  0.09}_{-  0.09}$ &	  1.44$_{-  0.05}^{+  0.08}\times10^{44}$\\
227 & 2.18 & 0.5 & 	 77.15$^{+ 46.04}_{- 16.24}$ &	  1.79$^{+  0.61}_{-  0.39}$ &	  1.15$_{-  0.53}^{+  1.38}\times10^{44}$\\
66 & 0.57 & 3.0 & 	  7.22$^{+  2.21}_{-  1.18}$ &	  1.59$^{+  0.44}_{-  0.19}$ &	  1.65$_{-  0.19}^{+  0.34}\times10^{43}$\\
67 & 1.62 & 3.0 & 	  0.00$^{+  0.62}_{-  0.00}$ &	  1.65$^{+  0.14}_{-  0.11}$ &	  1.13$_{-  0.07}^{+  0.07}\times10^{44}$\\
70 & 1.07 & 0.4 & 	 10.77$^{+  2.11}_{-  1.61}$ &	  1.40$^{+  0.09}_{-  0.00}$ &	  5.41$_{-  0.53}^{+  0.69}\times10^{43}$\\
4 & 1.26 & 1.6 & 	  0.00$^{+  0.57}_{-  0.00}$ &	  1.74$^{+  0.22}_{-  0.11}$ &	  4.59$_{-  0.46}^{+  0.28}\times10^{43}$\\
6 & 2.46 & 0.2 & 	  0.86$^{+  1.60}_{-  0.86}$ &	  1.89$^{+  0.21}_{-  0.17}$ &	  2.84$_{-  0.22}^{+  0.38}\times10^{44}$\\
11 & 2.58 & 3.0 & 	  0.00$^{+  0.74}_{-  0.00}$ &	  1.84$^{+  0.12}_{-  0.08}$ &	  4.84$_{-  0.24}^{+  0.30}\times10^{44}$\\
152 & 1.28 & 0.6 & 	 20.20$^{+  7.62}_{-  5.56}$ &	  1.81$^{+  0.59}_{-  0.41}$ &	  7.18$_{-  2.08}^{+  3.10}\times10^{43}$\\
159 & 3.30 & 0.5 & 	 10.14$^{+  4.47}_{-  3.98}$ &	  1.68$^{+  0.20}_{-  0.20}$ &	  4.52$_{-  0.79}^{+  1.10}\times10^{44}$\\
91 & 3.19 & 1.0 & 	  1.51$^{+  5.75}_{-  1.51}$ &	  1.40$^{+  0.31}_{-  0.00}$ &	  7.59$_{-  1.10}^{+  2.28}\times10^{43}$\\
24 & 3.61 & 3.0 & 	  1.12$^{+  5.77}_{-  1.12}$ &	  1.56$^{+  0.28}_{-  0.16}$ &	  2.57$_{-  0.29}^{+  0.83}\times10^{44}$\\
29 & 0.30 & 0.9 & 	  3.14$^{+  0.87}_{-  0.35}$ &	  1.55$^{+  0.36}_{-  0.15}$ &	  8.27$_{-  0.46}^{+  1.05}\times10^{42}$\\
265 & 1.22 & 1.5 & 	 15.71$^{+  9.94}_{-  4.76}$ &	  1.78$^{+  0.62}_{-  0.38}$ &	  2.91$_{-  0.85}^{+  2.35}\times10^{43}$\\
264 & 1.32 & 1.9 & 	 18.49$^{+ 19.80}_{-  6.83}$ &	  1.45$^{+  0.95}_{-  0.05}$ &	  1.45$_{-  0.39}^{+  1.99}\times10^{43}$\\
44 & 1.03 & 3.0 & 	  0.00$^{+  0.16}_{-  0.00}$ &	  2.26$^{+  0.11}_{-  0.06}$ &	  7.67$_{-  0.46}^{+  0.28}\times10^{43}$\\
45 & 2.29 & 1.5 & 	  9.44$^{+  5.16}_{-  3.39}$ &	  1.57$^{+  0.36}_{-  0.17}$ &	  1.20$_{-  0.18}^{+  0.38}\times10^{44}$\\
531 & 1.54 & 3.0 & 	 150 &	  1.80$^f$ &	  6.83$\times10^{43}$\\
156 & 1.19 & 3.0 & 	115.65$^{+ 35.11}_{- 29.02}$ &	  2.40$^{+  0.00}_{-  0.77}$ &	  2.18$_{-  1.30}^{+  0.82}\times10^{44}$\\
59 & 0.97 & 0.5 & 	  2.35$^{+  0.94}_{-  0.25}$ &	  1.78$^{+  0.27}_{-  0.26}$ &	  2.59$_{-  0.25}^{+  0.31}\times10^{43}$\\
61 & 2.02 & 0.5 & 	  1.10$^{+  0.92}_{-  0.88}$ &	  1.84$^{+  0.16}_{-  0.15}$ &	  2.42$_{-  0.19}^{+  0.22}\times10^{44}$\\
68 & 2.73 & 3.0 & 	  5.72$^{+  2.57}_{-  2.58}$ &	  1.96$^{+  0.20}_{-  0.21}$ &	  3.31$_{-  0.53}^{+  0.67}\times10^{44}$\\
72 & 1.99 & 0.5 & 	  6.82$^{+  2.26}_{-  2.00}$ &	  1.84$^{+  0.25}_{-  0.24}$ &	  1.80$_{-  0.25}^{+  0.33}\times10^{44}$\\
10 & 0.42 & 3.0 & 	  1.85$^{+  0.93}_{-  0.36}$ &	  1.40$^{+  0.31}_{-  0.00}$ &	  4.55$_{-  0.51}^{+  0.60}\times10^{42}$\\
15 & 1.23 & 1.5 & 	  0.10$^{+  0.65}_{-  0.10}$ &	  1.72$^{+  0.27}_{-  0.13}$ &	  3.89$_{-  0.33}^{+  0.40}\times10^{43}$\\
19 & 0.74 & 3.0 & 	  0.10$^{+  0.33}_{-  0.10}$ &	  1.82$^{+  0.23}_{-  0.14}$ &	  2.45$_{-  0.23}^{+  0.23}\times10^{43}$\\
146 & 2.67 & 0.5 & 	  1.92$^{+  1.25}_{-  0.95}$ &	  1.40$^{+  0.27}_{-  0.00}$ &	  6.62$_{-  1.27}^{+  1.49}\times10^{43}$\\
201 & 0.68 & 3.0 & 	  2.84$^{+  1.64}_{-  1.21}$ &	  1.85$^{+  0.55}_{-  0.45}$ &	  3.86$_{-  0.80}^{+  0.97}\times10^{42}$\\
41 & 0.67 & 3.0 & 	  7.30$^{+  2.98}_{-  1.69}$ &	  1.66$^{+  0.56}_{-  0.26}$ &	  1.61$_{-  0.26}^{+  0.47}\times10^{43}$\\
150 & 1.09 & 3.0 & 	 44.73$^{+ 19.89}_{- 13.23}$ &	  2.40$^{+  0.00}_{-  0.61}$ &	  3.99$_{-  1.55}^{+  1.35}\times10^{43}$\\
46 & 1.62 & 3.0 & 	  0.58$^{+  1.11}_{-  0.58}$ &	  2.04$^{+  0.34}_{-  0.21}$ &	  6.40$_{-  0.67}^{+  1.02}\times10^{43}$\\
47 & 0.73 & 3.0 & 	  8.63$^{+  2.90}_{-  3.16}$ &	  2.30$^{+  0.10}_{-  0.75}$ &	  1.04$_{-  0.33}^{+  0.36}\times10^{43}$\\
49 & 0.53 & 3.0 & 	  0.01$^{+  0.38}_{-  0.01}$ &	  1.70$^{+  0.34}_{-  0.18}$ &	  3.55$_{-  0.55}^{+  0.50}\times10^{42}$\\
263 & 3.66 & 3.0 & 	 77.22$^{+ 88.80}_{- 35.88}$ &	  1.41$^{+  0.99}_{-  0.01}$ &	  8.07$_{-  2.77}^{+ 32.89}\times10^{43}$\\
206 & 1.32 & 3.0 & 	  0.00$^{+  0.12}_{-  0.00}$ &	  2.06$^{+  0.08}_{-  0.08}$ &	  1.69$_{-  0.07}^{+  0.07}\times10^{44}$\\
57 & 2.56 & 3.0 & 	 18.31$^{+ 10.88}_{-  7.38}$ &	  1.65$^{+  0.42}_{-  0.25}$ &	  1.55$_{-  0.42}^{+  0.87}\times10^{44}$\\
58 & 0.92 & 0.5 & 	  4.16$^{+  1.12}_{-  1.69}$ &	  2.40$^{+  0.00}_{-  0.53}$ &	  1.01$_{-  0.22}^{+  0.22}\times10^{43}$\\
538 & 0.31 & 3.0 & 	  3.09$^{+  5.69}_{-  2.11}$ &	  1.40$^{+  1.00}_{-  0.00}$ &	  3.30$_{-  1.40}^{+  2.50}\times10^{41}$\\
114 & 1.72 & 0.5 & 	  2.66$^{+  2.69}_{-  1.85}$ &	  1.40$^{+  0.23}_{-  0.00}$ &	  2.71$_{-  0.49}^{+  0.53}\times10^{43}$\\
99 & 0.79 & 0.5 & 	  0.53$^{+  0.67}_{-  0.40}$ &	  1.50$^{+  0.29}_{-  0.10}$ &	  1.33$_{-  0.14}^{+  0.16}\times10^{43}$\\
73 & 0.73 & 3.0 & 	  0.55$^{+  0.53}_{-  0.46}$ &	  1.70$^{+  0.28}_{-  0.28}$ &	  9.41$_{-  1.00}^{+  1.12}\times10^{42}$\\
76 & 2.39 & 1.2 & 	 15.78$^{+  6.06}_{-  4.43}$ &	  1.63$^{+  0.35}_{-  0.22}$ &	  2.66$_{-  0.43}^{+  0.95}\times10^{44}$\\
208 & 0.72 & 0.6 & 	  0.53$^{+  0.62}_{-  0.53}$ &	  2.00$^{+  0.40}_{-  0.42}$ &	  4.79$_{-  0.80}^{+  1.04}\times10^{42}$\\
501 & 0.81 & 0.6 & 	  0.00$^{+  0.60}_{-  0.00}$ &	  1.73$^{+  0.24}_{-  0.14}$ &	  2.82$_{-  0.23}^{+  0.27}\times10^{43}$\\
12 & 0.25 & 3.0 & 	  0.00$^{+  0.05}_{-  0.00}$ &	  1.95$^{+  0.18}_{-  0.15}$ &	  7.10$_{-  1.10}^{+  1.10}\times10^{41}$\\
25 & 2.26 & 0.5 & 	 29.47$^{+  9.03}_{-  6.74}$ &	  1.40$^{+  0.09}_{-  0.00}$ &	  1.65$_{-  0.24}^{+  0.30}\times10^{44}$\\
609 & 1.86 & 0.5 & 	159.25$^{+273.38}_{- 56.16}$ &	  1.40$^{+  1.00}_{-  0.00}$ &	  5.03$_{-  0.66}^{+ 34.16}\times10^{43}$\\
190 & 0.73 & 3.0 & 	 13.89$^{+  7.62}_{-  3.65}$ &	  1.40$^{+  0.69}_{-  0.00}$ &	  1.02$_{-  0.25}^{+  0.46}\times10^{43}$\\
34 & 0.84 & 3.0 & 	  0.52$^{+  0.71}_{-  0.52}$ &	  1.58$^{+  0.32}_{-  0.18}$ &	  1.10$_{-  0.14}^{+  0.14}\times10^{43}$\\
203 & 1.17 & 0.7 & 	  0.61$^{+  0.85}_{-  0.60}$ &	  1.50$^{+  0.28}_{-  0.10}$ &	  4.47$_{-  0.41}^{+  0.52}\times10^{43}$\\
601 & 0.73 & 3.0 & 	255.19$^{+265.46}_{-118.40}$ &	  2.26$^{+  0.14}_{-  0.86}$ &	  7.39$_{-  5.83}^{+ 52.79}\times10^{43}$\\
83 & 1.76 & 0.5 & 	  0.00$^{+  0.92}_{-  0.00}$ &	  1.40$^{+  0.23}_{-  0.00}$ &	  3.43$_{-  0.39}^{+  0.42}\times10^{43}$\\
543 & 1.81 & 0.5 & 	 34.53$^{+ 21.09}_{- 12.40}$ &	  1.80$^{+  0.60}_{-  0.40}$ &	  8.76$_{-  3.85}^{+  8.37}\times10^{43}$\\
13 & 0.73 & 3.0 & 	  0.00$^{+  0.12}_{-  0.00}$ &	  1.83$^{+  0.18}_{-  0.13}$ &	  1.89$_{-  0.21}^{+  0.18}\times10^{43}$\\
200 & 0.85 & 0.4 & 	  0.82$^{+  1.12}_{-  0.52}$ &	  1.40$^{+  0.30}_{-  0.00}$ &	  8.02$_{-  1.18}^{+  1.60}\times10^{42}$\\
30 & 0.84 & 3.0 & 	  0.00$^{+  0.43}_{-  0.00}$ &	  1.87$^{+  0.28}_{-  0.15}$ &	  4.80$_{-  0.55}^{+  0.51}\times10^{43}$\\
35 & 1.51 & 3.0 & 	  8.36$^{+  4.09}_{-  4.13}$ &	  2.05$^{+  0.35}_{-  0.55}$ &	  2.21$_{-  0.68}^{+  0.66}\times10^{44}$\\
48 & 1.26 & 0.5 & 	  3.20$^{+  2.08}_{-  0.95}$ &	  1.40$^{+  0.34}_{-  0.00}$ &	  2.92$_{-  0.40}^{+  0.57}\times10^{43}$\\
54 & 2.56 & 3.0 & 	 11.97$^{+ 10.69}_{-  4.83}$ &	  1.53$^{+  0.58}_{-  0.13}$ &	  9.34$_{-  1.87}^{+  5.84}\times10^{43}$\\
97 & 0.18 & 2.0 & 	  0.00$^{+  0.07}_{-  0.00}$ &	  1.41$^{+  0.23}_{-  0.01}$ &	  4.60$_{-  0.90}^{+  0.40}\times10^{41}$\\
207 & 0.40 & 0.4 & 	  5.57$^{+  1.01}_{-  1.69}$ &	  2.40$^{+  0.00}_{-  0.68}$ &	  6.14$_{-  1.25}^{+  1.13}\times10^{43}$\\
65 & 1.10 & 0.5 & 	  1.21$^{+  1.09}_{-  1.01}$ &	  1.96$^{+  0.44}_{-  0.41}$ &	  1.65$_{-  0.26}^{+  0.25}\times10^{43}$\\
71 & 1.04 & 3.0 & 	  0.00$^{+  0.44}_{-  0.00}$ &	  1.72$^{+  0.19}_{-  0.12}$ &	  3.34$_{-  0.29}^{+  0.28}\times10^{43}$\\
75 & 0.74 & 3.0 & 	  4.71$^{+  2.82}_{-  1.00}$ &	  1.41$^{+  0.58}_{-  0.01}$ &	  3.01$_{-  0.48}^{+  0.75}\times10^{43}$\\
7 & 1.84 & 0.6 & 	  0.68$^{+  3.48}_{-  0.60}$ &	  1.70$^{+  0.35}_{-  0.12}$ &	  5.04$_{-  0.82}^{+  1.90}\times10^{44}$\\
23 & 0.73 & 0.5 & 	  0.00$^{+  0.45}_{-  0.00}$ &	  1.84$^{+  0.33}_{-  0.19}$ &	  5.79$_{-  0.84}^{+  0.90}\times10^{42}$\\
243 & 2.50 & 0.3 & 	 16.66$^{+ 14.60}_{-  8.74}$ &	  1.40$^{+  0.64}_{-  0.00}$ &	  6.91$_{-  2.22}^{+  5.50}\times10^{43}$\\
267 & 0.72 & 1.2 & 	  9.24$^{+  7.20}_{-  2.67}$ &	  1.40$^{+  0.81}_{-  0.00}$ &	  9.36$_{-  2.64}^{+  5.30}\times10^{42}$\\
213 & 0.60 & 0.5 & 	  1.94$^{+  8.09}_{-  1.27}$ &	  1.42$^{+  0.98}_{-  0.02}$ &	  4.41$_{-  3.64}^{+  8.45}\times10^{42}$\\
209 & 1.32 & 0.5 & 	  1.26$^{+  0.88}_{-  1.03}$ &	  1.55$^{+  0.33}_{-  0.15}$ &	  9.91$_{-  1.29}^{+  1.13}\times10^{43}$\\
503 & 0.54 & 0.4 & 	  0.35$^{+  0.37}_{-  0.33}$ &	  1.87$^{+  0.31}_{-  0.29}$ &	  8.32$_{-  1.14}^{+  1.04}\times10^{42}$\\
\enddata\\
\tablecomments{The table lists in turn XID, redshift and its quality (see Zheng et al. 
2004), best fit N$_H$, powerlaw photon index $\Gamma$ and 2 -- 10 keV rest frame absorption 
corrected luminosity. Following T06, 0.6 -- 7 keV band spectra are adopted
(binned to at least 1 photon per bin) for spectral fit, and C-statistic 
(Cash 1979) for minimization.
All the quoted errors are $90\%$ confidence range 
for one parameter of interest. The primary model is an absorbed 
powerlaw with Galactic absorption included. The uncertainties of luminosities 
are estimated by varying $\Gamma$, N$_H$ and powerlaw normalization to include
parameter sets with $\Delta$C $<$ 2.706. 
The varying range of $\Gamma$ is set to 1.4 to 2.4 (see text for details).
Sources with N$_H$ fixed at 
1.5 $\times$ 
10$^{24}$ are better fitted (with $\Delta$C $<$ - 4) by the reflection model 
($pexrav$) and its intrinsic luminosities are estimated assuming the reflected
luminosity in 2 -- 10 keV band is about 6\% of the intrinsic one (see T06 for
details for spectral models and fitting procedures).
}
\end{deluxetable}

\clearpage

\begin{references}  
\reference{}Alexander, D. M., et al. 2003, AJ, 126, 539
\reference{}Antonucci, R. 1993, ARA\&A, 31, 473
\reference{}Barger, A. J., et al. 2005, ApJ, 129, 578
\reference{}Brandt, W. N., et al. 2001, AJ, 122, 2810
\reference{}Caccianiga, A., et a. 2004, A\&A, 416, 901
\reference{}Cash, W. 1979, \apj, 228, 939
\reference{}Comastri, A., Fiore, F., Vignali, C., Matt, G., Perola, G. C. \& La Franca, F. 2001, MNRAS, 327, 781
\reference{}Dickey, L. M., \& Lockman, F. J. 1990, ARA\&A, 28, 215
\reference{}Dwelly, T. et al. 2005, MNRAS, 360, 1426
\reference{}Dwelly, T. \& Page, M. J. 2006, MNRAS in press, astro-ph0608479
\reference{}Eckart, M. E. et al. 2006, ApJ in press, astro-ph0603556
\reference{}Fiore, F. et al. 1999, MNRAS, 306, L55
\reference{}Fiore, F. et al. 2003, A\&A, 409, 79
\reference{}Freeman, P. E., Kashyap, V., Rosner, R., Lamb, D. Q. 2002, ApJS, 138, 185
\reference{}Geheres, N, 1986, ApJ, 303, 336
\reference{}Georgantopoulo, I., Georgakakis, A. \& Akylas, A. 2006, A\&A accepted, astro-ph/0610828
\reference{}Giacconi, R., et al. 2001, ApJ, 511, 624
\reference{}Giacconi, R., et al. 2002, ApJS, 139, 369
\reference{}Kormendy, J., \& Gebhardt, K. 2001, in AIP Conf. Proc. 586, 20th Texas Symposium on Relativistic Astrophysics, ed. H. Martel \& J. C. Wheeler (Melville: AIP), 363
\reference{}Kormendy, J., \& Richstone, D. 1995, ARA\&A, 33, 581
\reference{}La Franca, F. et al. 2005, ApJ, 635, 864
\reference{}Mainieri, V. et al. 2002, A\&A, 393, 425
\reference{}Mart\'\i nez-Sansigre, A. et al. 2005, $Nature$, 436, 666
\reference{}Nandra, K., Georgantopoulos, I., Ptak, A., \& Turner, T. J. 2003, ApJ, 582, 615
\reference{}Norman, C. A. et al. 2002, ApJ, 571, 218
\reference{}Perola, G. C. et al. 2004, A\&A, 421, 491
\reference{}Piconcelli, E., Cappi, M., Bassani, L., Fiore, F., Di Cocco, G., Stephen, J. B. 2002, A\&A 394, 835
\reference{}Piconcelli, E., Cappi, M., Bassani, L., Di Cocco, G., Dadina, M. 2003, A\&A, 412, 689
\reference{}Rosati, P. et al. 2002, ApJ, 566, 667
\reference{}Spergel, D. N. et al. 2003, ApJS, 148, 175
\reference{}Steffen, A. T., Barger, A. J., Cowie, L. L., Mushotzky, R. F., Yang, Y. 2003, ApJ, 596, L23
\reference{}Stern, D. et al. 2002, ApJ, 568, 71
\reference{}Szokoly, G. P., et al. 2004, ApJS, 155, 271
\reference{}Tozzi, P, et al. 2001, ApJ, 562, 42
\reference{}Tozzi, P, et al. 2006, A\&A, 451, 457, T06
\reference{}Ueda, Y., Akiyama, M., Ohta, K., Miyaji, T. 2003, ApJ, 598, 886
\reference{}Yu, Q. \& Tremaine, S. 2002, MNRAS, 336, 965
\reference{}Zheng, W. et al. 2004, ApJS, 155, 73
\end{references}
\end{document}